\documentclass[11pt]{cernrep}
\usepackage[pdftex]{graphicx,epsfig}
\usepackage{amsmath,amssymb}
\usepackage[dvips,letterpaper,text={6.5in,9in}]{geometry}
\usepackage{fancyhdr}
\usepackage{verbatim}
\usepackage{color}
\usepackage{floatrow}
\usepackage{blindtext}
\usepackage{url}

\newcommand\ltap{\
  \raise.3ex\hbox{$<$\kern-.75em\lower1ex\hbox{$\sim$}}\ }
\newcommand\gtap{\
  \raise.3ex\hbox{$>$\kern-.75em\lower1ex\hbox{$\sim$}}\ }
 
 \renewcommand{\phi}{\varphi}

\newcommand\simge{\mathrel{%
   \rlap{\raise 0.511ex \hbox{$>$}}{\lower 0.511ex \hbox{$\sim$}}}}
\newcommand\simle{\mathrel{
   \rlap{\raise 0.511ex \hbox{$<$}}{\lower 0.511ex \hbox{$\sim$}}}}

\newcommand{\slashchar}[1]%
        {\kern .25em\raise.18ex\hbox{$/$}\kern-.75em #1}
\def\lsim{\mathrel{\raise.3ex\hbox{$<$\kern-.75em\lower1ex\hbox{$\sim$}}}}
\def\gsim{\mathrel{\raise.3ex\hbox{$>$\kern-.75em\lower1ex\hbox{$\sim$}}}}

\begin{document}
\title{\includegraphics[scale=0.95]{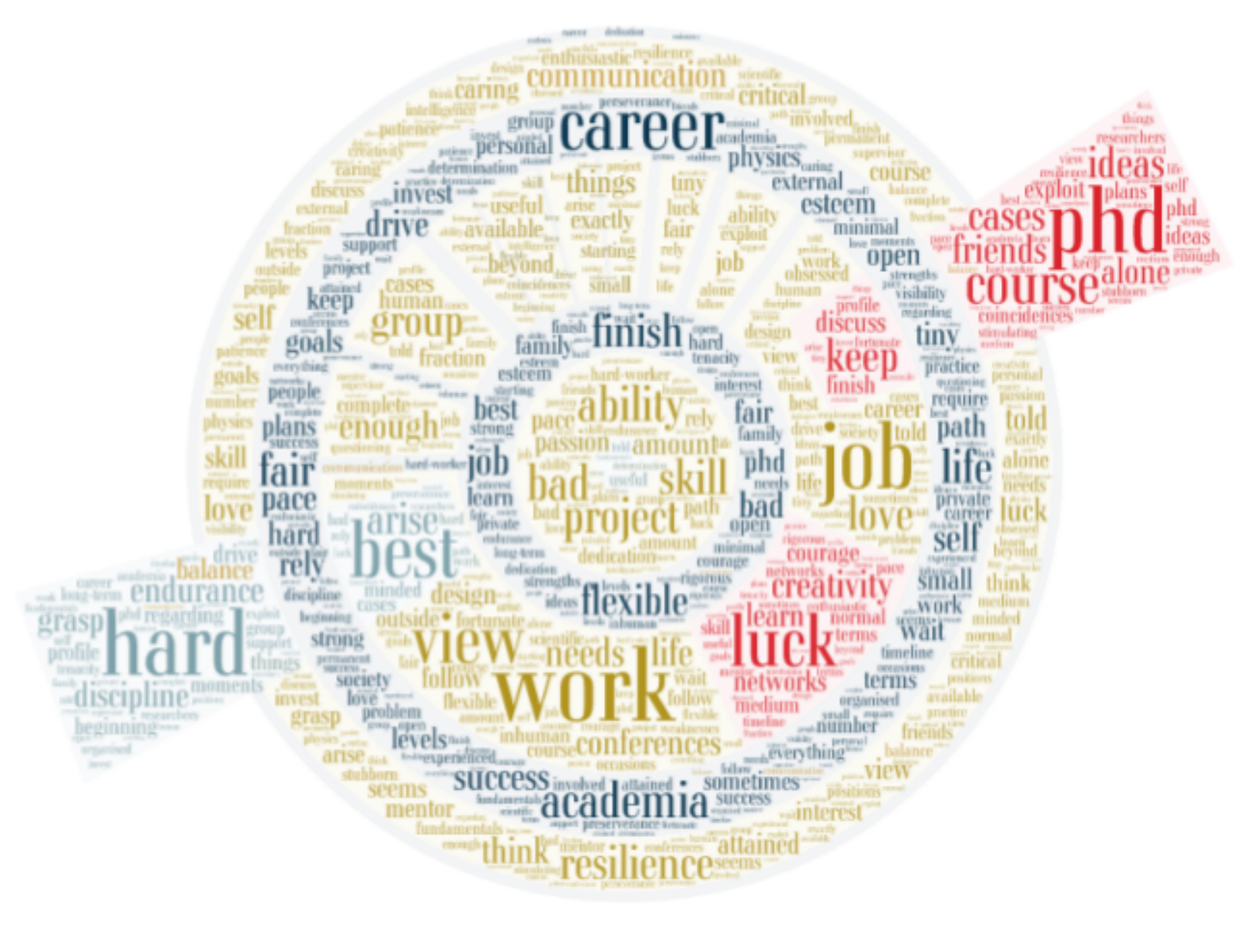}\\ Developing Careers in Physics {--} Perspectives of Particle Physics Researchers from the VBScan network at various stages of their careers\\VBSCAN-PUB-09-19}

\author{Kristin~Lohwasser$^1$}
\institute{
$^1$Department of Physics and Astronomy, Sheffield University, Sheffield, UK }

\maketitle

\begin{abstract}
Outlooks of particle physics researchers on their careers and the general challenges in establishing their careers over different career stages are surveyed using a questionnaire distributed to participants in an ERC-funded research network, ``VBScan''. The respondents displayed a great deal of insight into what is needed for a career in academia, or ore specifically particle physics, though they also did not downplay the element of ``luck''. Some notable differences between career levels could be observed in problems raised and attitudes towards careers.
\end{abstract}


\newpage

\section{Introduction}
\label{sec:intro}

Navigating an (academic) career or just understanding one's role in the wider community of one's field of research can be a difficult experience. Internationally, about 20\% of PhD students will stay in academia or continue working in fundamental research~\cite{dpgstudy,belgiumstudy}. This insecurity before tenure as well as the different tasks and duties of their roles, the challenges faced by PhD students, post-doctoral researchers, junior and senior staff members can be quite different. Studies show that career goals and perceptions change over time due to the increased influence of personal relationships and disillusionment with career prospects in the field of study~\cite{doi:10.1080/03075079.2014.914908,10.1371/journal.pone.0036307}. 

Particle physics and particle phycistics have been the object of anthropological and sociological studies since its beginning~\cite{beamlines,quarks,culture}, being an ideal playground to study decision making and construction of scientific truth or knowledge~\cite{doi:10.14506/ca29.3.03}. More recently, the focus has shifted towards the impact that large collaborations have on crediting of scientific work~\cite{Birnholtz} and more specifically on authorship~\cite{doi:10.1080/08989621.2014.968277,Pritychenko:2015dma} and also knowledge acquisition inside the collaboration~\cite{Camporesi2017}. That having a career in particle physics research and that there are implicit requirements and expectations, that are not always made transparent has already been pointed out by Sharon Traweek in Ref.~\cite{beamlines}: 
\begin{quotation} It is widely understood among the senior group members in American labs that it is not sufficient for a postdoc to do this sort of mundane job well; independent, risky work must be undertaken as well.  So far as I know, no one tells this to the postdoc. Some sense it immediately; some discover it when they realize that their hard labor is not getting the attention they want; others never learn.
\end{quotation}

Whilst further work describes very briefly the perceived means of researchers trying to distinguish themselves within a large collaboration~\cite{Birnholtz}, it does little to explore the outlooks of these researchers on their careers and their possible struggles in establishing their careers over the different stages of their career. The present work aims to close this gap using a structured questionnaire which explores first general attitudes towards physics and the initial interest in the subject before exploring attitudes towards careers, what it takes to have a career in physics and how they judged their place in physics changed over the years and how researchers felt about their current position within the field.

\section{Methodology and Sample Properties}
\label{sec:method}

The questionnaire was created to be online accessible and a link was distributed via various VBScan\footnote{The main goal of the VBSCan project is to investigate the Vector Boson Scattering (VBS) process and its implications for the Standard Model, by coordinating existing theoretical and experimental efforts in the area and by best exploiting hadron colliders data, thereby laying the groundwork for long-term studies of the subject and creating a solidly interconnected community of VBS experts.  (\url{https://vbscanaction.web.cern.ch/})} network mailing lists (totalling about 150 subscribers). The VBScan network is a scientific network sponsored by the European Research Council (ERC) within their COST network programme\footnote{\url{https://www.cost.eu/}}. It is open to researchers around the globe, but travel funds are generally only available to members of the participating European research institutions and meetings are organised only in European member countries\footnote{The 38 COST Member Countries are: Albania, Austria, Belgium, Bosnia and Herzegovina, Bulgaria, Croatia, Cyprus, Czech Republic, Denmark, Estonia, Finland, France, Germany, Greece, Hungary, Iceland, Ireland, Italy, Latvia, Lithuania, Luxembourg, Malta, the Republic of Moldova, Montenegro, The Netherlands, The Republic of North Macedonia, Norway, Poland, Portugal, Romania, Serbia, Slovakia, Slovenia, Spain, Sweden, Switzerland, Turkey, United Kingdom. Non-COST Members can join based on mutual benefit, they are spread across the Near Neighbour Countries and International Partner Countries.}. In general, the questions were formulated to be relatively open and required free text answers, that allowed respondents to expand on those aspects that they were most interested in, providing a qualitative overview over attitudes towards particle physics and careers.

\begin{figure}[h!]
\centering
\includegraphics[width=0.6\textwidth]{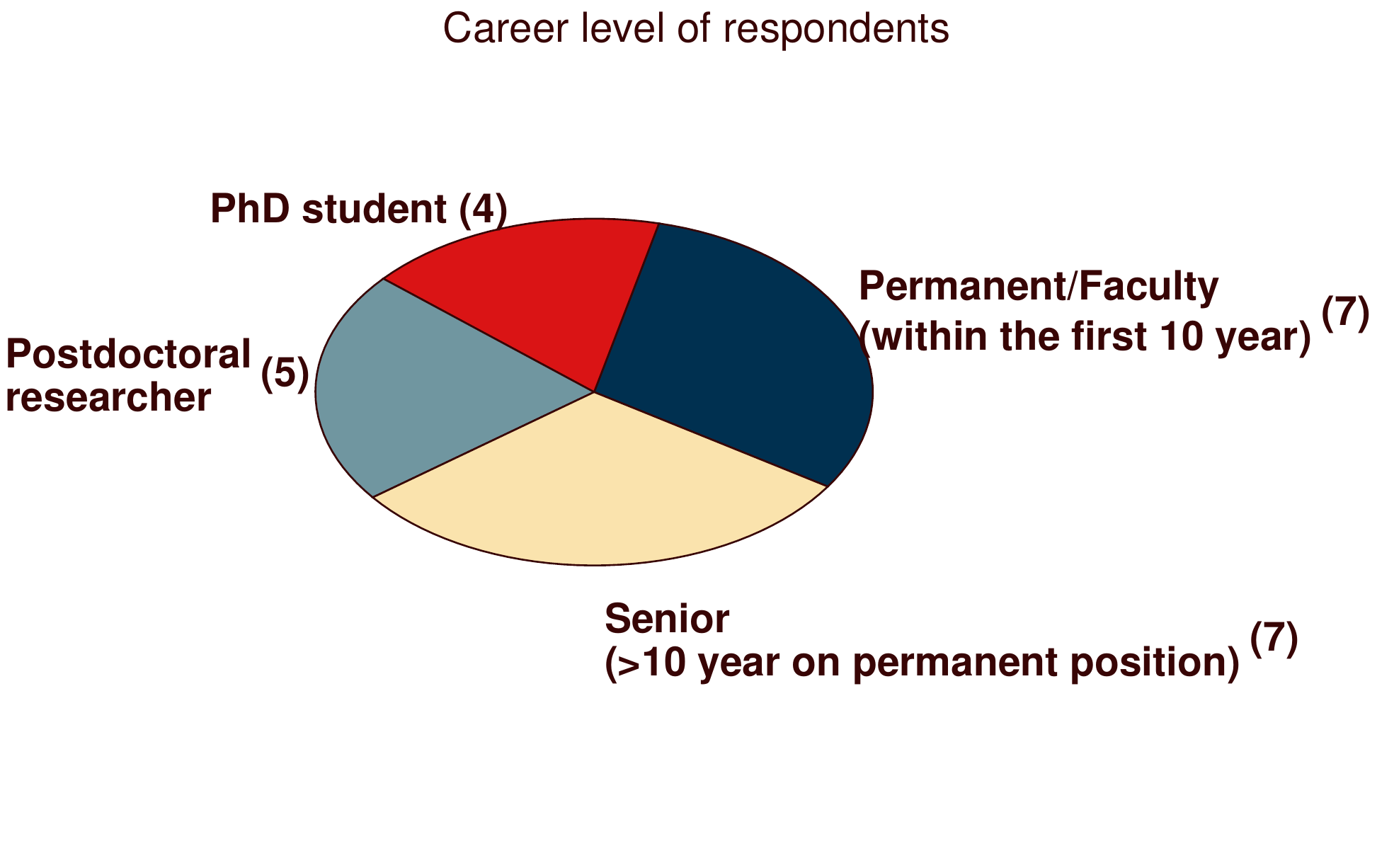}
\caption{
\label{fig:careerlevels} Career levels of the respondents: It is split up roughly equally between permanent (60\%) and non-permanent researchers (PhDs, post-docs, 40\%).}
\end{figure}

Out of the more than 150 possible respondents, 23 replied. Out of those, 12 prefered to stay anonymous (the questionnaire gave the option to fill in an email address), whilst 11 people allowed themselves to be identifiable. Fig.~\ref{fig:careerlevels} summarizes the career levels of the respondents. Slightly less PhD students (4 or 17.4\%) and post-doctoral researchers (5 or 21.7\%) have replied compared to permanent researchers (7 or 30.4\% for both junior and senior researchers). Still, numbers are similar for the different career levels. In general however, answers from post-doctoral researchers, junior and senior permanent researchers were much longer and generally more detailed than those of PhD students. Five respondents were female, corresponding to 21.7\%, which is reflective of the number of females in the field~\cite{ATL-GEN-PUB-2016-001}.

\begin{figure}[h!]
\centering
\includegraphics[width=0.8\textwidth]{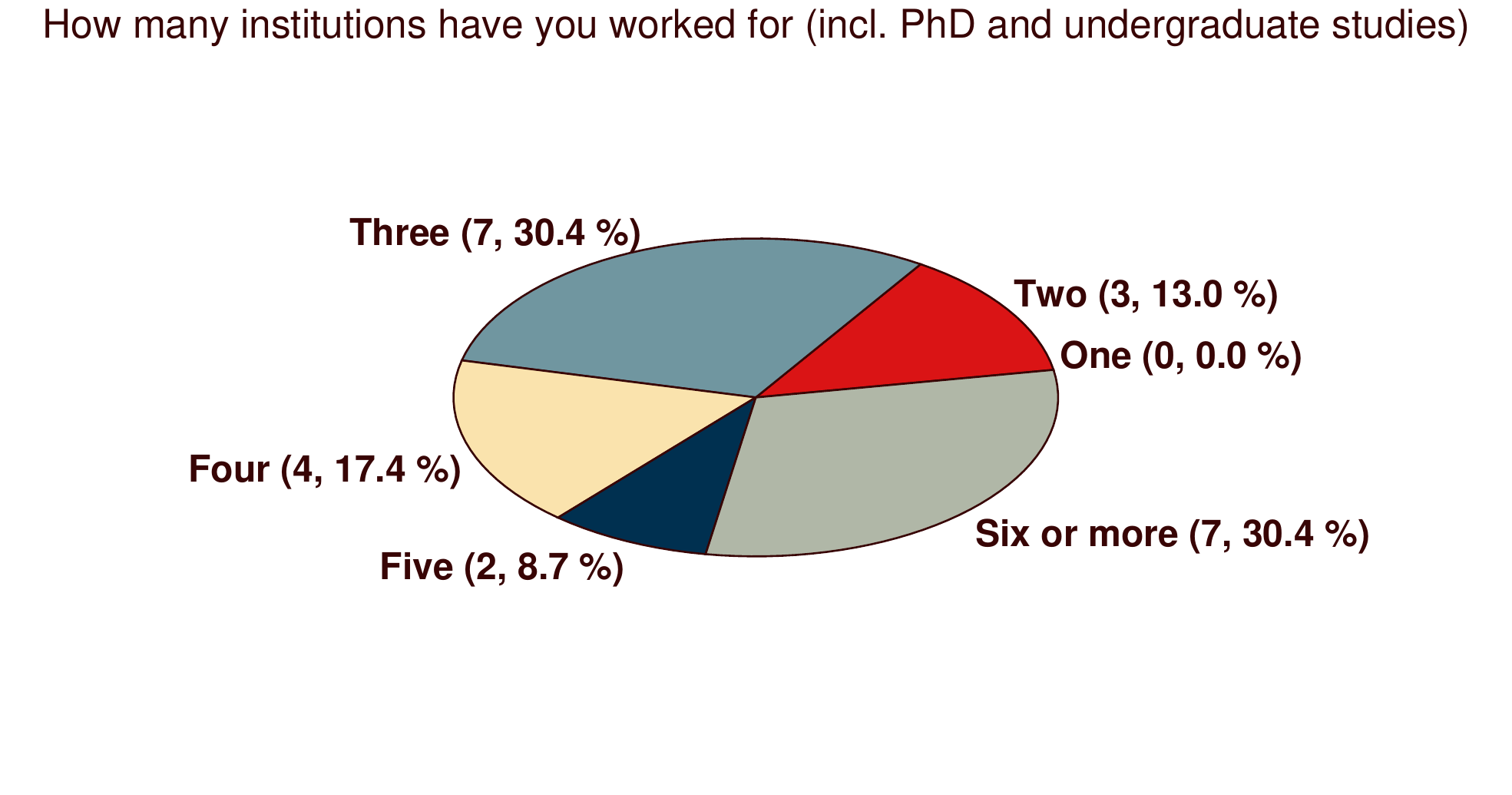}
\caption{
\label{fig:numberofinstitutes} Number of institutes respondents studied at and worked for.}
\end{figure}

The geographical origin (``country of origin'') of participants reflects roughly the dominance of researchers from Italy, Germany, France and Spain in the VBScan network (5:4:3:3). Further three respondents came from other European countries and one respondent originated from outside of Europe. As probably to be expected in the very international environment of particle physics, about a third of respondents (8) were not living in their country of origin. Also a third had studied or completed their PhD somewhere other than their country of origin (8). Five (21.7\%) were still working abroad after they had done their PhD abroad, six (26.1\%) did either work abroad or had done their PhD abroad, but not both. As a consequence of the high mobility of the researchers, everyone had worked for more than one institution. 30.4\% had already already worked at six or more institutions as summarised in Fig.~\ref{fig:numberofinstitutes}. All of the 4 PhD students that had answered had changed institute for their PhD. 

\section{General attitude towards Physics}
\label{sec:physics}

Before concentrating on career advice and career goals in a more narrow sense, the questionnaire aimed to investigate, what drew people towards a career in physics (or specifically particle physics) in the first place and how they felt their initial fascination or understanding of physics had changed over the years.

\subsection{Inspiration: Why physics?}

\textit{What did inspire you to study physics? What was the thing you most loved when studying physics? Why did you decide to pursue a career in physics / do a PhD? }

Despite the fact that the answers were formulated freely, a few main themes emerged. Doing physics and research is to do with challenge, problem solving, understanding nature and how things work. People expressed a genuine interest, often as a motivation to study physics with the specific interest in particle physics being sparked at university.    

Most used in the answers were the following words (frequency given in brackets, removing such words as particle, physics, liked, love, work, decided): challenge (3) curiosity (3) interest (4) mathematics (3) nature (5) problems (4) research (4) study (4) understanding (5) university (3) how things work (4). 

A few people cited specific popular science shows or books that sparked their interest: \textit{Dexter’s Lab, Stephen Hawkings, Mr Tompkins (Gamow), Time life magazines about the big bang \& particle reactions in the early universe}. 

Two quite typical answers given are:

\begin{quotation}
My secondary school teacher inspired me to study physics. I liked most to solve difficult problems. I did a PhD in physics because I couldn't get enough difficult problems. [Postdoctoral Researcher]
\end{quotation}

\begin{quotation}
Most loved in my studies were stages of increasing level of understanding about complex matters; doing research is just like doing cross-words, but for much more complicated and much cooler problems. [Permanent/Faculty (within the first 10 year)]
\end{quotation}

\subsection{Change throughout the years}

Asking for the difference between expectations and how physics changed over the years for people\footnote{\textit{And how were things different from what you expected / how did physics *change* for your over the years?}}, answers became a bit less idealistic and show some disillusionment:

\begin{quotation}It is full of stress. We are always in a rush to catch a non-realistic deadline which is imposed by someone who is not working in the analysis and it feels like running with no end. I would really like it if we had more time. Also, I started my PhD when there was a lot of excitement around the community in the field and as the years pass, this has lessened a lot. [PhD student]
\end{quotation}

A reference to \textit{formalities and visibilty} by another student was echoed by a postdoctoral researcher, who described the university to be \textit{no different than the corporate world}. This is also reflected to some extend by a senior researcher pointing out that \textit{[t]he personal initiative, autonomous learning and the ability of defining one personal scientific and professional plan turned out to be more important than what [they had] initially perceived during University studies}. 

A few people at postdoctoral or junior researcher level mentioned that things got increasingly difficult or complicated and complex, refering both to the actual research problems as well as to the environment (in terms of how research is organised). Most senior researchers mentioned administration and increased bureaucracy as obstacles. Only two people out of the 19 respondents reported an overall better experience than expected. 

A question more focused on the positive aspects followed\footnote{\textit{What do you enjoy most about your current job / your current "role in physics"?}}, that invited people not to dwell on unexpected problems as possibly the previous one, but to concentrate on the perks of the jobs. 

There were a lot of commonalities between the answers of the 21 respondents, most of them independent of their career stage. More than half (11) enjoyed \textit{collaboration} and working with (many different) people, with more senior researchers emphasizing the working with young people and generally mentioning teaching in addition. The second-most mentioned point is \textit{freedom} or \textit{flexibility} (7, 33.3\%), of either schedules or of organising one's own work or or pursuing one's interests. ''Learning new things every day`` was mentioned by 5 (23.8\%) of respondents, in addition thinking about new ideas or working creatively appears twice as answer (9.5\%).  

\section{Careers in Physics}
\label{sec:career}
 
The questionnaire further aimed to explore, both what people thought was expected for a career from the wider community as well as what they thought they needed as advice. Already in Refs.~\cite{beamlines,Birnholtz} a few points were made: \textit{dependable, diligent, responsible, and willing to work long hours, to come up with novel solutions to difficult problems (analysis, detector design or construction problem).}

Indeed a few of them were also made in answer to \textit{What do you think it necessary for a successful PhD or a career in physics/academia?}\footnote{The word cloud created from all answers to this question has been added on the title page.}. At least 8 people mentioned \textit{hard-work} or working \textit{beyond the time of a normal 9-to-5 job}, and further 5 variations appearead in the answers such as \textit{stamina} or \textit{perseverance} (\textit{Endurance and perseverance in once's work, both in good and bad moments.}). A PhD student mentioned \textit{Hard work and inhuman levels of motivation}, indicating that this attitude is indeed still prevailent. 

Communication skills as well as \textit{scientific networking, presence and visibility at workshops and conferences} were mentioned 8 times (38\%) as well, emphasizing the need for collaboration inside the particle physics community, which has not been highlighted that much specifically in older works. Notable is however, that especially younger people emphasized networking and communication. This might indicate that either their importance has changed (due to the increased size of the collaboration), or that it is so ingrained for the more senior people that it is not so much noteworthy.

Beyond collaboration, support from a mentor or supervisor but also from friends and family is recognized (3 or 14\%): \textit{Trying to keep a good balance between work and private life.} Further two respondents added resilience to that list with one of them saying: 
\begin{quotation}Partly creativity and curiosity, but also a lot of resilience, on occasions low self esteem, being a white man is also a useful skill. [Postdoctoral researcher]
\end{quotation}

This is the only response, that discusses specific challenges for females, although this remark here is unspecific, specific challenges have been studied and discussed in Ref.~\cite{Gurney1985}.  

Three respondents quoted luck as an external factor (\textit{And a little bit of luck (from the human and the job-market point of view)}) as an important ingredient in a job market as tight as academia. 

A PhD student gave the most comprehensive description, including external and internal factors: 

\begin{quotation}
Regarding external factors, 1. a caring supervisor, 2. a concrete timeline for the phd from the beginning (even though it seems to me that this cannot happen, because it depends on the analysis group goals and pace, and also new things that arise, but I think it would be nice to have one), and 3. a group to discuss. 
Regarding the personal work, questioning everything you learn and do, and this is how new ideas come. Present your work often, do not wait until you finish the whole project. Also, try to be involved in many things, as many as you can, to be as more "complete" as possible. [PhD student] 
\end{quotation}

Asked about \textit{the most common mistakes} people made in their career planing or when working in physics, often similar things were relayed, namely about visibility and work-life balance (\textit{[it is a mistake to just have] physics-related life and friends (hanging out only with colleagues, "living" in the office, etc)}).

A few new points were brought up: 
\begin{quotation}
[A] lack of modesty and of optimism can be an issue during interviews. [Permanent/Faculty (within the first 10 year)]
\end{quotation}
As well as: 
\begin{quotation}
assuming that just because you are good you will get the job. [another Permanent/Faculty (within the first 10 year)]
\end{quotation}

Most mentioned however was \textit{focus} (4 times mentioned verbatim) and related topics such as lack of \textit{long-term goals} and \textit{getting distracted }or giving up on something. This included getting too focused on one issue (\textit{tunnel vision}) as well as not balancing \textit{between personal interest and community interests}. In summary, lack of consistency and lack of focus on long- and middle-term achievements aligned with the interested of the broader field were identified as issues or mistakes. 

Interestingly, the more senior the respondents, the less concrete the advice on how to achieve a career or what mistakes to avoid becomes, often just a few words such as \textit{Endurance, resilience, passion and determination} were given. 

Asked, whether they (still) wanted a career\footnote{\textit{Would you like to have a career in physics (if you already have one: would you like to do it under the current circumstances?)? Why? Why not?}}, more than half (13) said unambiguously \textit{yes}, whilst 3 gave \textit{maybe} as an answer and 4 did not respond. One person said outright \textit{no}: 
\begin{quotation}
When I look at "successful" colleagues (full professors, heads of department) I think I wouldn't like to be in their place in 20 years from now. [Postdoctoral Researcher]
\end{quotation}

For some people the \textit{yes} seems well motivated: 
\begin{quotation}
I like working in physics because I am mostly free to select which work I do; moreover, the work I should do is usually well-motivated and not forced down my throat (I worked outside of academia for some time) [Postdoctoral Researcher]
\end{quotation}

Those who were not sure, were mostly permanent researchers, who stated:
\begin{quotation}
Maybe, too much time wasted on secondary issues (but no reference wrt industry from own experience) [Permanent/Faculty (within the first 10 year)]

I just started my career in physics and there are aspects, that I am still not sure about/sold too. Who knows, I might change career at some stage to work less or to move to another place where I want to live [another Permanent/Faculty (within the first 10 year)]

there's still a lot of fun in it, but I got somewhat desillusioned along the way [Senior ($>$10 year on permanent position)]
\end{quotation}

\section{Career advice}
\label{sec:careeradvice}

A further set of questions was asked to understand, what career advice people would like to have for their own career or their own current worklife\footnote{\textit{What are the things, that you personally would like to have advise on for your career/your worklife?}. Separate answers were requested for each career level and it was stressed, that also advice from PhD students to e.g. senior researchers was welcome (such that every person should be able to answer all four advice questions for all four career levels): \textit{This could be *also* advice from a PhD to a senior staff member (e.g. "please, do not give us as much guidance... / please, give us more guidance... )}.}. Naturally, answers were very different for the different career stages. 

PhD students were most interested into the transition between PhD to postdoctoral researchers, with questions evolving around how to choose the best next step: 
\begin{quotation}
How to chose the best workplace for me. Seems that there are opportunities but how to choose the one that serves my future best. Choosing by the subject of the postdoc, the institute, the group, the postdoc on which you develop new skills? [PhD student]
\end{quotation}

Postdoctoral researchers were more occupied with the question of how to get a permanent position, how to apply for grants or even \textit{on how to secure private funding for research (i.e. through agreements with tech companies, performing consulting activities for them, part time job etc.)}. Another point raised was the question of uniqueness and how to demonstrate it. 

Junior permanent researchers were interested in questions of balance, between research and science politics/grant/networking, or of the general workload (possibly referring to work-life balance). Management issues as well as interest in a \textit{global overview of problems as seen by more senior people} were also mentioned.

There were very few responses from senior researchers (especially given, they usually provided very reflected and detailed answers compare to e.g. PhD students). The only ones were \textit{Be optimistic} and \textit{Stay away of your duties and do what you like}. 

Further, specific advice was asked for the different career levels, in an attempt to investigate, what people thought to be most helpful for a specific career stage\footnote{\textit{What is advice that you can give / would like to give? You could give an example of a situation where this advice would have helped}}. It turned out, that these indeed were much more specific answers than those aimed to figure out what to do to have a career. 

Concerning the specific advise for PhDs, postdoctoral researchers concentrated on their well-being (\textit{One day after another, talk to people when you get stuck, don't neglect yourself, make sure you are happy for at least one day a week, keep going}, \textit{do not be scared of building your own life/family... it is a hard life, but it is doable}) and only gave very general work advice (\textit{Work on something to understand it}, \textit{Work regularly}). Advice to PhD students from junior and senior permanent researchers centered on them broadening their horizons and indeed for them to try to to gain independence (\textit{try to follow your ideas}, \textit{explore your interests, also beyond your supervisors interests}, \textit{keep questioning what you are [beging told]}). 

\begin{quotation}
Talk to your supervisor. Tell her/him what works, what works well, and what doesn't work (well). Feedback in both directions is important. Try to be selective with your topic, and with the place you do your Phd. Look for places and supervisors with impact, but chemistry between supervisor/student is also important. Try to find a mentor, someone at the level of a permanent researcher or a postdoc other than your supervisor. Like the research you do, otherwise it is not the right one. There will be phases of frustration, keep going. [Permanent/Faculty (within the first 10 year)]
\end{quotation}

\begin{quotation}
try to read a lot of different stuff and see, what you can understand. Talk to people, listen to what the senior people discuss - this will broaden your knowledge beyond what you are actually doing, which is very very useful. Try to be proactive (even if it's not perfect), but try to get advise/feedback early on your ideas. [Permanent/Faculty (within the first 10 year)]
\end{quotation}
 
Advice for postdoctoral researchers was basically the same on life and work from Phds and postdoctoral researchers themselves (\textit{do not be scared of building your own life/family... it is a hard life, but it is doable}), whilst there was a pronounced shift in the advice from for postdocs from permanent researchers. It's obvious anyhow, but also clear from the answer, that the postdoc phase is deemed to be the decisive one, \textit{the most difficult phase. Scientifically the best and most proliferatic, but the most uncertain one, with many changes}.

Similar to the advice to students, independence was praised (\textit{Insist on having freedom to assign time according to your decision}), as well as setting and working towards goal. However, for postdocs, \textit{[learning] skills in grant writing and networking} was emphasized as well as to further expanding experiences: 
\begin{quotation}
Find time to engage in teaching. It gives a broader view of our field and can lead to fruitful collaboration with students. Strive to win your own research funding to show you can lead independent research. [Permanent/Faculty (within the first 10 year)]
\end{quotation}

The decisiveness of the postdoc career stage was underlined by advice to also think about when to end it {--} independently of the outcome:
\begin{quotation}
Put in a balance if you want to pay the price of being wandering around until 30's or 40's, specially if you have a family. Set an exit time cut-off. Do not extend it for ever. [Permanent/Faculty (within the first 10 year)]
\end{quotation}

Advice for junior staff researchers was only given by junior and senior researchers, however with very different focus. The advice from junior to junior staff was focused on finding a new balance in that role, along the lines of the following advice:

\begin{quotation}
The most important thing is to find the right balance between partner/family and career. This applies already to earlier stages. For junior staff, one has to split the work time between research, grant work, sometimes teaching and on the other hand supervising the members of the group. This is a difficult task. Try to find a good balance here. This should be weighted by the demands of a tenure procedure, or the demands of finding a permanent position (which is mostly research output) if the junior position is fixed-term. [Permanent/Faculty (within the first 10 year)]
\end{quotation}

Senior staff members mainly advised to develop one's research portfolio (\textit{Develop a unique feature in your work}, \textit{Don't stop. Keep going. Don't follow beaten paths, explore new direction}) and did not comment on other aspects of academic careers, academia in general or life. 

On the other hand, there was much more advise given by PhDs, postdoctoral and junior researchers to senior researchers. Mainly these are wishes on how senior researchers should fill out their role, namely \textit{teach[ing] skills like grant writing}, \textit{[setting a] timeline and clear goals from the start of the PhD} and \textit{[giving] guidance to PhD students}. Seniors themselves mainly concentrated on how to treat colleagues and collaborators, which also includes students: \textit{Show responsibility for your collaborators}, \textit{Be nice to young colleagues and motivate them}. Common between both groups of research is the appeal to be open to change and acknowledge the need to evolve with it:

\begin{quotation}
Be open to changes in the way how research, teaching, human relations, networking, coding, etc. is evolving in these years. [Permanent/Faculty (within the first 10 year)]
\end{quotation}

\begin{quotation}
Stay young, don't do more of the same. [Senior ($>$10 year on permanent position)]
\end{quotation}

\section{Conclusions}
\label{sec:conclusions}

As in other publications~\cite{wisskarriere} almost all respondents describe a career in science to be highly desirable, however almost all seemed to be acutely aware of the difficulties and the unpredictability of attaining a permanent position in research and academia. This has at least the advantange of the non-permanent researchers being aware and on some level being more pro-active in developing a career, which should be useful also outside academia. It is evident, that also the more senior researchers are aware of the problems of the academic system, which is also reflected in the strong appeal to take care of collaborators. 

There are some further indications, that career perceptions change over time though for the respondents of the study this seems to correlate most with roles that are changing over time and an increased responsibility in teaching and adminstration. It is noteable that with age, the downsides of academia (work-life-balance) and the high competition are much less often verbalised. It is hard to tell, whether this has to do with settling down in a permanent position in academia specifically or whether this is more generally a ''coming of age`` and a development of serenity that has calming effects. This question could presumably be answered if also people having left academia could be included in a follow-up study.

\section*{Acknowledgments}

Thanks to the respondents of the questionnaire and the support through the COST Action CA16108. K.L. is further supported by the European Union's Horizon 2020 research and innovation programme under ERC grant agreement No. 715871.

\bibliography{CareerSurvey}
\bibliographystyle{utcaps}
\end{document}